\documentclass[%
reprint,
 amsmath,amssymb,
 aps,
longbibliography,
]{revtex4-1}

\usepackage{graphicx}
\usepackage{grffile} 
\usepackage{subfigure}
\usepackage[marginal]{footmisc}
\usepackage{epstopdf}
\usepackage{amssymb}
\usepackage{mathrsfs}
\usepackage{amsmath,mathtools,amsfonts,bm,graphicx}
\usepackage{upgreek}
\usepackage{color}
\usepackage{etoolbox}
\usepackage{multirow}
\usepackage[normalem]{ulem}
\usepackage{enumitem}

\newcommand{\sect}[1]{Sec.~\ref{#1}}
\newcommand{\Sect}[1]{Section~\ref{#1}}
\newcommand{\fig}[1]{Fig.~\ref{#1}}
\newcommand{\Fig}[1]{Figure~\ref{#1}}
\newcommand{\eq}[1]{Eq.~(\ref{#1})}
\newcommand{\eqr}[2]{Eqs.~(\ref{#1})-(\ref{#2})}
\newcommand{\eqs}[2]{Eqs.~(\ref{#1}) and (\ref{#2})}
\newcommand{\Eq}[1]{Equation~(\ref{#1})}

\newcommand{\App}[1]{Appendix~\ref{#1}}

\newcommand{\rev}[1]{{#1}}

\newcommand{\av}[1]{\left\langle #1 \right\rangle}

\newcommand{\hd}[1]{{\bf  {#1:}~}}

\newcommand{\w}{w_0}
\renewcommand{\S}{\Phi}
\newcommand{\Sij}{\S_{ij}}

\newcommand{\M}{\mathcal{M}}

\begin{document}

\title{Emergent facilitation by random constraints in a facilitated random walk model of glass}
\author{Leo S.I. Lam$^1$}
\author{Hai-Yao Deng$^2$} 
\author{Wei-Bing Zhang$^3$}
\author{Udoka Nwankwo$^1$}
\author{Chu Xiao$^4$}
\author{Cho-Tung Yip$^4$} 
\author{Chun-Shing Lee$^4$} \email[Email: ]{anakin-cs@hotmail.com}
\author{Haihui Ruan$^1$} \email[Email: ]{haihui.ruan@polyu.edu.hk}
\author{Chi-Hang Lam$^5$} \email[Email: ]{C.H.Lam@polyu.edu.hk}
\address{
$^1$Department of Mechanical Engineering, Hong Kong Polytechnic University, Hong Kong, China\\
$^2$School of Physics and Astronomy, Cardiff University, 5 The Parade, Cardiff CF24 3AA, Wales, UK\\
$^3$School of Physics and Electronic Sciences, Changsha University of Science and Technology, Changsha 410004, China.\\
$^4$Department of Physics, Harbin Institute of Technology, Shenzhen 518055, China\\
$^5$Department of Applied Physics, Hong Kong Polytechnic University, Hong Kong, China
}

\date{\today}

\begin{abstract}
  The physics of glass has been a significant topic of interest for decades. Dynamical facilitation is widely believed to be an important characteristic of glassy dynamics, but the precise mechanism is still under debate. We propose a lattice model of glass called the facilitated random walk (FRW). Each particle performs continuous time random walk in the presence of its own random local kinetic constraints. The particles do not interact energetically. Instead, they interact kinetically with a hopping rate resampling rule under which motions of a particle can randomly perturb the local kinetic constraints of other particles. This dynamic interaction is reversible, following a rate restoration rule.
  A step-by-step reversal of the particle motions exactly restore the previous constraints, modeling randomness quenched in the configuration space of glass. The model exhibits  stretched exponential relaxation and dynamical heterogeneity typical of glasses. Despite the lack of explicit facilitation rule, the FRW shows facilitation behaviors closely analogous to those of the kinetically constrained  models (KCM).  The FRW is  a coarse-grained version of the distinguishable particle lattice model (DPLM) and this  exemplifies that compatible defect and atomistic models can complement each other on the study of glass.
\end{abstract}

\maketitle

\section{Introduction}
\label{intro}

The understanding of the dominant modes of relaxation dynamics in glassy materials has been debated actively for decades \cite{biroli2013review,arceri2022review,wang2019review,rodriguez2022review,zaccone2023review,dyre2024review,yu2024review}. Molecular dynamics (MD) simulations are instrumental in their studies as they provide complete information on the atomistic details of motions \cite{barrat2010review,barrat2023review}. Besides more realistic all-atom simulations, simplified coarse-grained models are widely believed to capture essential properties of glass. Nevertheless, due to the extremely slow dynamics of glass, long-time relaxation dynamics in MD simulations at deep supercooling relevant to experimental conditions are still computationally inaccessible, despite the availability of equilibrium samples using advanced swap algorithms.  The dominant relaxation dynamics of glass can be hidden behind vibrational and other liquid-like collective motions which diminish at deeper supercooling.

Lattice models play pivotal roles in many branches of statistical physics \cite{plischke1994,barabasi1995}. Many phenomena can be qualitatively reproduced and certain characteristic quantities, such as scaling exponents, are in agreement with values from off-lattice simulations and experiments. Many lattice models for glass are successful in reproducing key features such as kinetic arrest and dynamical heterogeneity \cite{garrahan2011review,ritort2003review}. There are however great challenges to justify assumptions in most of these models. In addition, many glassy features may not be readily reproduced. 

The $n$-spin facilitated model proposed by Fredrickson and Andersen (FA) \cite{fredrickson1984,fredrickson1985}  is one of the first kinetically constrained models (KCM)
\cite{garrahan2011review}. It is a coarse-grained defect model of glass. An up-spin represents a defect in the form of a local region with low particle density. The flipping of a spin can only be facilitated by the presence of at least $n$ neighboring up-spins.
The dynamical facilitation picture pioneered by the FA and related models   \cite{fredrickson1984,fredrickson1985,palmer1984,chandler2010review,chandler2011} has been gaining further support very recently \cite{scalliet2022,costigliola2024}.

There have been proposals of new lattice models or variants  attempting to better capture the essence of glass \cite{zhang2017,tahaei2023,ozawa2023,hasyim2024,nishikawa2024}. In particular, the distinguishable particle lattice model (DPLM) \cite{zhang2017} of glass   is defined by particle pair interactions and void-induced dynamics. The DPLM is an atomistic model in which each particle represents an atom or a rigidly bounded group of atoms. Voids correspond to missing particles \cite{cangialosi2011,yip2020,mei2022}.
\rev{They are simplified representations of quasivoids found in glassy colloidal liquids \cite{yip2020} as motivated by glassy bead-spring polymer simulations \cite{lam2017} and have also been suggested previously as free-volume holes in polymer experiments \cite{cangialosi2011}. The physical relevance of quasivoids to glass in general, however, is yet to be studied.} The DPLM demonstrates emergent facilitation behaviors and has reproduced an extraordinarily wide range of kinetic and thermodynamic characteristics of glass \cite{zhang2017,lulli2020,lee2020,lulli2021,lee2021,gao2022,gopinath2022,gao2023,li2023,ong2024,zhai2024}.

In this paper, we propose a facilitated random walk (FRW) model of glass. It is a  coarse-grained and energetically trivialized version of the DPLM. It is a defect model with particles corresponding to voids in glass \cite{yip2020,cangialosi2011,mei2022}. The FRW may open up a new class of KCM with random constraints, which can be more readily justifiable than the usual deterministic kinetic constraint rules \cite{garrahan2011review}. We will explain emergent facilitation behaviors in which dynamics are dominated by coupled particle groups with a group size depending on the constraint density. The facilitated dynamics in such mobile groups are analogous to  the facilitation rule in the FA model. 

The FRW model is a generalization of simpler random walk models which will first be summarized. Additional rules specific to FRW will be explained in the next section. Only one-dimensional (1D) systems will be discussed but generalization to higher dimensions is straightforward.

\hd{Continuous-time random walk (CTRW)}
Consider $N$ particles on a 1D lattice  with $L$ sites under periodic boundary conditions. 
No exclusion rule is imposed so that any site $i$ can contain $n_i = 0, 1, 2, \dots$ particles.
The particle density is  $\rho = N/L$, which can be varied from 0 to $\infty$.

For simple CTRW with no kinetic constraint, every particle can hop to a given nearest neighboring (NN) site with a rate $w_0$. With two hopping pathways to the two NN sites in 1D, the combined hopping rate of a particle to an arbitrary NN site is thus $2w_0$.

\hd{CTRW with quenched constraints}
We now impose a distinct set of local kinetic constraints for each particle by randomly blocking some of the hopping pathways. Specifically, particle $k$ hops from site $i$ to a NN site $j$ with a rate $w_{ijk}$, given by $w_{ijk}=w_0$ with an unblocking probability $q$ and $w_{ijk}=0$ otherwise. We put $w_{ijk}=w_{jik}$ in order to follow detailed balance. Otherwise, each $w_{ijk}$ is an independent and quenched random variable sampled at the beginning of a simulation.
For $q=1$, simple CTRW is restored. For $q<1$, every particle is locally trapped within a finite well of its own. 
With no exclusion principle nor energetic interaction,
particle motions are independent of each other and are trivially solvable.
\rev{Randomness in the constraints is quenched in the real space, i.e., the random constraints follow a time-invariant function of position for each particle and do not evolve when particle configuration changes. This type of disorder is more akin to that for spin glass \cite{mezard1990book} and is, however, not appropriate for glass.}

The rest of the paper is organized as follows. In \sect{model}, we will generalize the simple models above to the FRW model. \Sect{equil} explains its exact equilibrium statistics and the kinetic Monte Carlo simulation algorithm used. In \sect{glass_char}, we demonstrate standard glassy behaviors of FRW including stretched exponential relaxations. \Sect{facil} explains intuitively the emergent facilitation mechanism and the resulting ergodic property. We then conclude in \sect{conclusion} with further discussions.

\section{Model}
\label{model}

\begin{figure}[tb]
  \includegraphics[width=0.9\columnwidth]{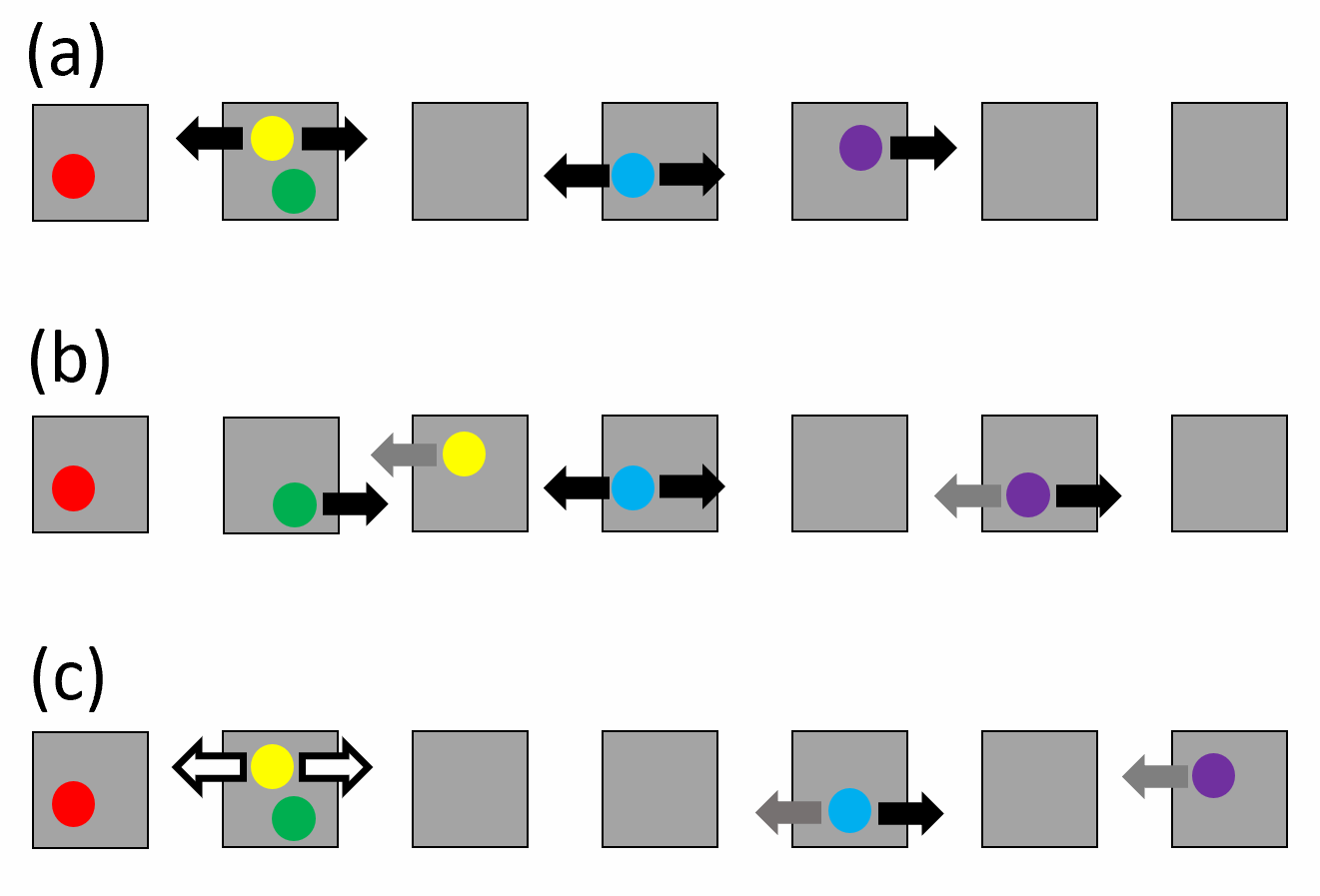}
  \caption{
A schematic showing an example of system evolution in FRW for a lattice of $L=8$ sites (grey squares) with $N=5$ particles (colored circles).
\rev{An arrow (black, grey or open), occurring with an unblocking probability $q=0.5$, indicates that the particle is able to hop at rate $w_0$ in the indicated direction.
During the period shown, the red particle is trapped, while the blue and purple particles move independently of other particles.
Starting from the initial configuration (a), the hop of the yellow particle has led to a rate resampling of the right-hopping rate of the green particle (b). In (c), the reversed hop of the yellow particle has restored the previous hopping rates of itself (open arrows) and of the green particle.
In (b)-(c), possible hops indicated by grey arrows must be present due to detailed balance.}
}
\label{diagram}
\end{figure}

The FRW model is a coarse-grained defect model of glass, similar to the FA model \cite{fredrickson1984}. Each site represents a mesoscopic region with, for example,  dozens of particles. A particle in FRW physically represents a defect in the form of a void \cite{yip2020}. Atoms or molecular units in the glass are not explicitly simulated. All results are averaged over at least two simulations.

Specifically, the model in 1D is constructed based on CTRW with random constraints in the absence of particle exclusion.
Similar to the descriptions in \sect{intro}, we consider a lattice size $L$, lattice constant $a=1$, total number of particles $N$, density $\rho=N/L$, and occupancy $n_i=0, 1, 2, \dots$ at site $i$.
As before, the random constraints are such that the  
hopping rate $w_{ijk}(t)$ of particle $k$ from site $i$ to a NN site $j$ depends on the unblocking probability $q$ and follows 
\begin{equation}
  \label{wqt}
  w_{ijk}(t) = 
  \begin{cases}
  w_0 & ~~~~~ \mbox{probability} ~~ q\\
  0 &  ~~~~~ \mbox{otherwise.}
  \end{cases}
\end{equation}
with $w_0= 1$ and they obey the detailed balance condition
\begin{equation}
  \label{detailedbalance}
   w_{ijk}(t) = w_{jik}(t).
 \end{equation}
For the FRW model, we emphasize, however, that the hopping rate $w_{ijk}(t)$ is $not$ quenched but can be reversibly resampled, i.e. resampled or restored, repeatedly throughout a simulation. The rules for this time dependence are the defining characteristics of FRW and are defined  as follows:
\begin{enumerate}[label=(\roman*)]
\item {\bf Rate resampling:} If particle $k$ hops from site $i$ to $j$, we will resample the hopping rate $w_{ijl}(t)=w_{jil}(t)$ across bond $ij$ for all other particles $l\neq k$ using \eq{wqt}.
\item {\bf Rate restoration:} After particle $k$ has hopped from site $i$ to $j$, if it performs a reversed hop from $j$ to $i$, we will restore the previous values of $w_{ijl}(t)=w_{jil}(t)$ across bond $ij$ for all particles $l\neq k$.
\end{enumerate}
Figure \ref{diagram} shows an example of particle configuration and evolution.

The resampling rule introduces a form of dynamical interactions, in the absence of any energetic interaction.
It turns the FRW into a correlated-particle model and enables rich dynamical behaviors.
Physically, it models the mechanism that the motion of particle  $k$, representing a defect, perturbs the local packing of the glass atoms. This alters the energetically favorable local hopping pathways of every other defect $l$. However, the effect  is felt only when $l$ arrives at the perturbed sites. 

On the other hand, the rate restoration rule implements randomness quenched in the configuration space but not in the real space. It physically represents that when defect $k$ reverses its previous hop, the local atomic packing of the glass is reversed so that the previous local energetically favorable pathways of every other defect $l$ are restored.

In general, the restoration rule can be applied to any arbitrarily long sequence of hops. If the hops are reversed step-by-step in the reversed order, the rates will be step-by-step restored. Nevertheless, in a large system, groups of particles that are well separated at all times should not affect each other.  More precisely, the restoration rule applies to a whole correlated sequence of hops. It does not require reversing any other uncorrelated hop at a distance which has no impact on the rates of hops in the sequence.

\rev{The present model does not involve any temperature explicitly. Nevertheless,
the effects of temperature can be introduced via the unblocking probability $q$ and the particle density $\rho$. One expects
that as temperature decreases, both $q$ and $\rho$ decrease.}

\section{Equilibrium states and simulation algorithm}
\label{equil}
The FRW is a kinetic model with trivial energetics. The detailed balance condition (\eq{detailedbalance}) implies that all accessible states have identical occurrence probability in an equilibrium ensemble, implying the same system energy at all states. Furthermore, the model is ergodic in the large system size limit, as will be explained in \sect{facil}. As a result, the equilibrium states of FRW is trivial and can be described exactly as follows. 
At equilibrium, all particle arrangements are possible and are equally likely.
To initialize an equilibrium system, we simply need to put each particle randomly and independently on the lattice.

We perform FRW simulations using a rejection free Monte Carlo algorithm. During a simulation, we continuously keep track of all unblocked hopping directions  with rate $w_0$ of all particles. At every time step, we randomly choose one of these hops with a uniform probability. To choose efficiently, we
search a binary tree which dynamically represents all available hops.  
Then, the hop is performed and the simulation time advances by $(Mw_0)^{-1}$ where $M$ is instantaneous value of the total number of unblocked hopping directions of all particles. 

A potentially challenging part of the algorithm is related to the hopping rate restoration rule, which recalls previous rates after an arbitrarily long hopping sequence is step-by-step reversed.
To enable such restoration, one could employ a memory intensive algorithm that involves saving the history of all previous states and hopping rates into the computer memory. However, this method requires a nontrivial data structure design and prohibitively large amount of memory for long runs.

To solve this rate restoration problem, we have developed a reversible pseudo random number based algorithm so that all required rates are calculated from random number sequences. The random number generators thus  encode and effectively store all possible rates. Only the current system state needs to be directly stored in the computer memory.

The idea of the rate restoration algorithm is as follows. The FRW system state is primarily characterized by the positions $i_k$ of every particle $k$. For bookkeeping purpose, we introduce fictitious internal states $\Psi_k$ and $\Sij$ of particle $k$ and bond $ij$. These internal states are represented by 64-bits random numbers so that no two states will be accidentally equal in practical simulations.
A hop involves updating $\Psi_k$ and $\Sij$ essentially to the next numbers in the pseudo random number sequences. At a reversed hop, previous states are  obtained by updating to the previous numbers in the sequences, which are readily calculable because we adopt  reversible generators. The hopping rate $w_{ijk}(t)$ is defined as  a function of the instantaneous states $\Psi_k$ and $\Sij$ so that previous rates can be readily restored once the previous states $\Psi_k$ and $\Sij$ are restored.
Using this memory-less algorithm, we are able to run the simulation for long time without any memory concerns while maintaining exactly the rate restoration property.
This algorithm is explained in detail in Appendix \ref{states} and  \ref{algo}. 
The main simulation codes based on these algorithms is available at \cite{leogithub}.

We have extensively tested the reliability of our software implementation. In particular, we have checked that the known equilibrium states are arrived at and maintained throughout the kinetic Monte Carlo simulations. Furthermore, a highly non-trivial test is the quantitative verification of an exact mobility threshold for two-particle systems, as will be reported elsewhere.

\section{Glassy characteristics}
\label{glass_char}
We now explain FRW simulation results, demonstrating typical glassy characteristics.

\begin{figure}[tb]
  \includegraphics[width=0.9\columnwidth]{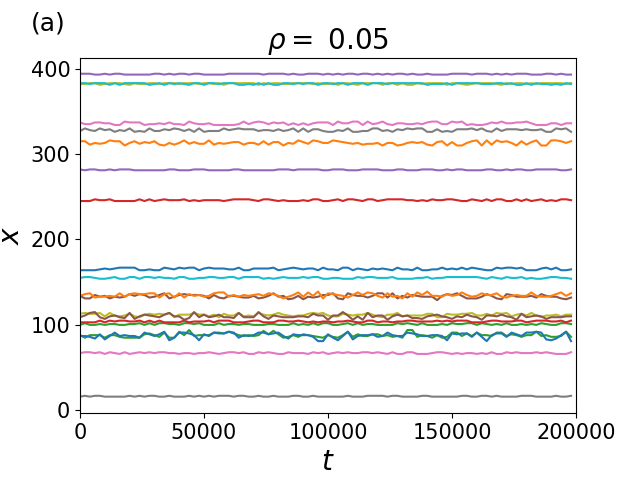}
  \includegraphics[width=0.9\columnwidth]{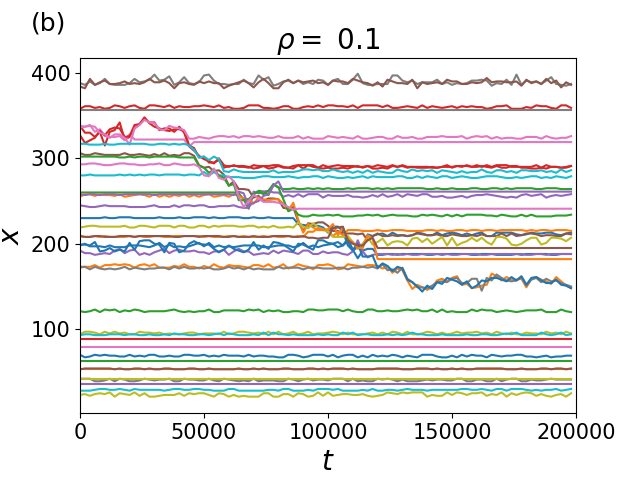}
  \includegraphics[width=0.9\columnwidth]{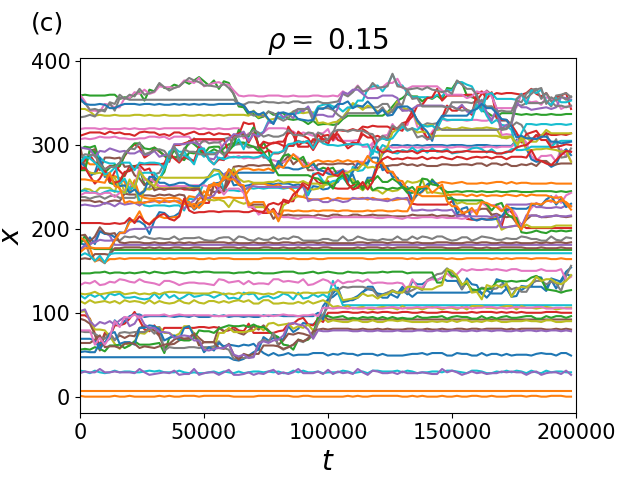}
  \caption{Position-time graphs of particles for lattice size $L=400$, unblocking probability $q=0.5$ and density $\rho=0.05$ (a), 0.1 (b)  and $0.15$ (c). Mobile groups in (b) and (c)  each consist of 3 particles most of the time. The lack of any mobile group in (a) is only a finite size effect.}
  \label{traj}
\end{figure}

\begin{figure}[tb]
  \includegraphics[width=0.9\columnwidth]{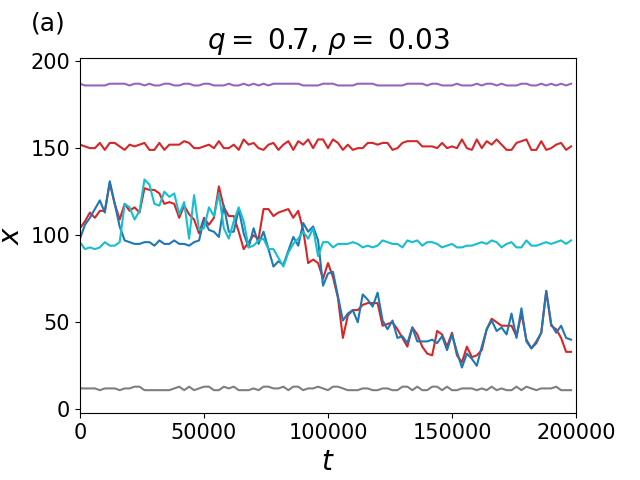}
  \includegraphics[width=0.9\columnwidth]{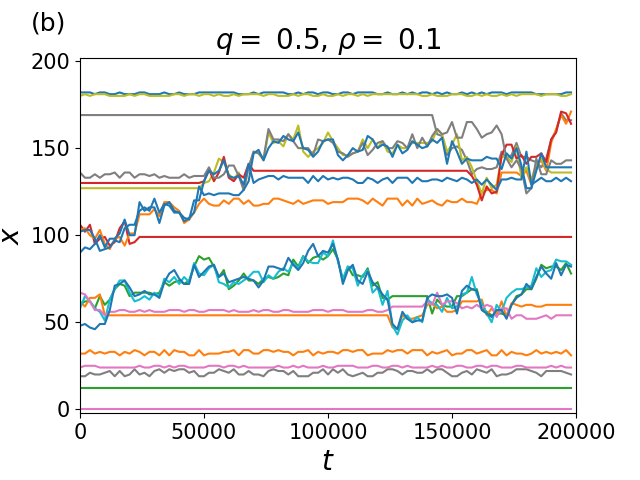}
  \includegraphics[width=0.9\columnwidth]{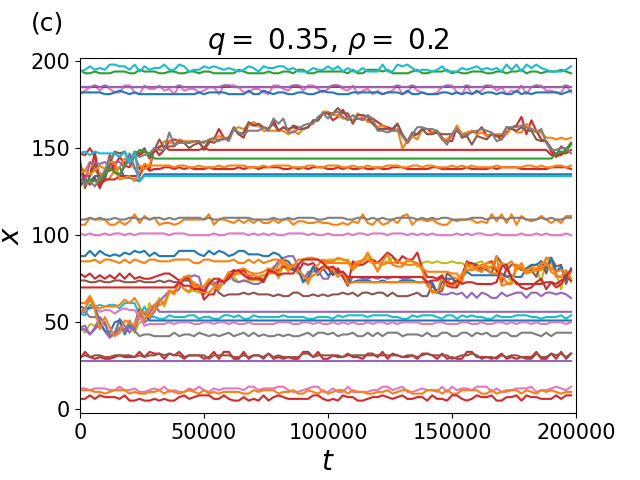}
  \caption{Position-time graphs of particles for lattice size $L=200$, unblocking probability $q=0.7$ (a),  0.5 (b) and 0.35 (c). To reveal relatively isolated mobile groups, we take density $\rho=0.03$ (a),  0.1 (b) and 0.2 (c). Each mobile group usually  consists of 2, 3, and 4 particles in (a), (b) and (c) respectively.}
  \label{traj_q}
\end{figure}

\hd{Dynamical facilitation}
Position-time graphs in \fig{traj} illustrate particle motions in equilibrium FRW systems for an  
unblocking probability $q=0.5$ under different particle density $\rho$. In \fig{traj}(a) with a low density $\rho=0.05$, particles are mostly isolated and they are all permanently trapped within small regions. Coupled pairs of particles have movements confined to slightly wider traps. We will explain in \sect{facil} that such permanent trapping is only a finite size effect.

In \fig{traj}(b) where the density is doubled to $\rho=0.1$, a mobile group, usually consisting of 3 particles, traverses over much of the system and can indeed travel unboundedly at long time. As the mobile group moves, it picks up, drops off or exchanges particles while keeping at least 3 particles at all times. When not part of the mobile group, particles are generally trapped at their respective locations. Importantly, the drastically higher mobility observed when particles are coupled together demonstrates an emergent facilitation phenomenon which will be intuitively explained in \sect{facil}.

Figure \ref{traj}(c) corresponding to an even higher density $\rho=0.15$ shows more abundant mobile groups. Each group also contains at least 3 particles. Group sizes fluctuate due the close proximity to many other mobile or trapped particles. 

The dominant size of mobile groups in FRW decreases with the unblocking probability $q$. To show this, we have selected values of $q$ where the typical mobile group sizes are 2, 3 and 4 as plotted in \fig{traj_q}(a), (b) and (c) respectively. A different $\rho$ is used in each case to give only one or two mobile groups for clarity. This dependence is explained in \sect{facil}.

\hd{Plateaus in Mean Square Displacement}
\begin{figure}[tb]
  \includegraphics[width=0.9\columnwidth]{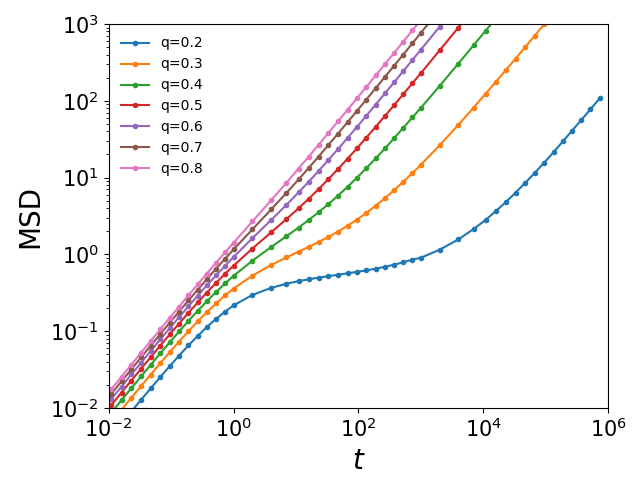}
  \caption{Particle MSD against time $t$ in log-log scales for  various unblocking probability $q$ and density $\rho = 0.8$.}
  \label{msd}
\end{figure}
We calculate the particle mean square displacement (MSD) defined as
\begin{equation}
  MSD = \av{\left\vert x_k(t)-x_k(0)\right\vert^2}
  \label{MSD}
\end{equation}  
where $x_k(t)$ denotes the  position of particle $k$ at time $t$. \Fig{msd} shows the MSD against time for different $q$ and $\rho=0.8$. At long time, the slopes of the lines in the log-log plot is close to unity, indicating the diffusive regime. Subdiffusive plateaus, characteristic of glass, appear at intermediate time at low $q$. They indicate temporary trapping of particles either isolated or in small immobile groups as described above.
At long time, most momentarily trapped particles have been repeatedly picked up and displaced by the larger mobile groups, resulting at the diffusive regime.
We observe that plateaus are more pronounced at small $q$. 
This reveals stronger particle trapping and more glassy at small $q$. More extensive results also show that a smaller $\rho$ also increases the glassiness.
MSD for $\rho=0.2$ showing qualitatively similar features is presented in \App{rho02}.

\hd{Self-intermediate scattering function}
\begin{figure}[tb]
  \includegraphics[width=0.9\columnwidth]{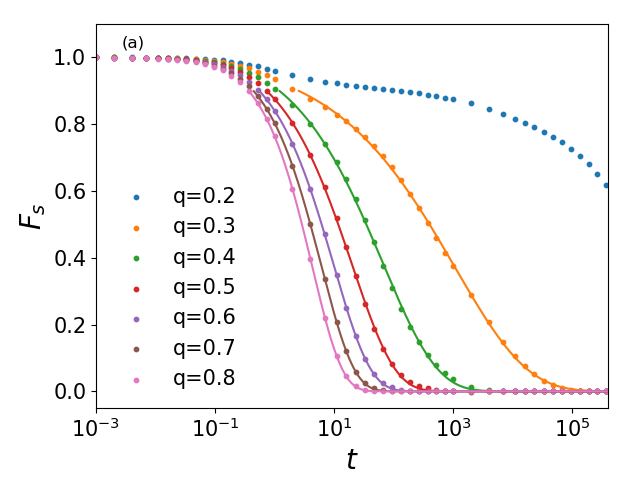}
  \includegraphics[width=0.9\columnwidth]{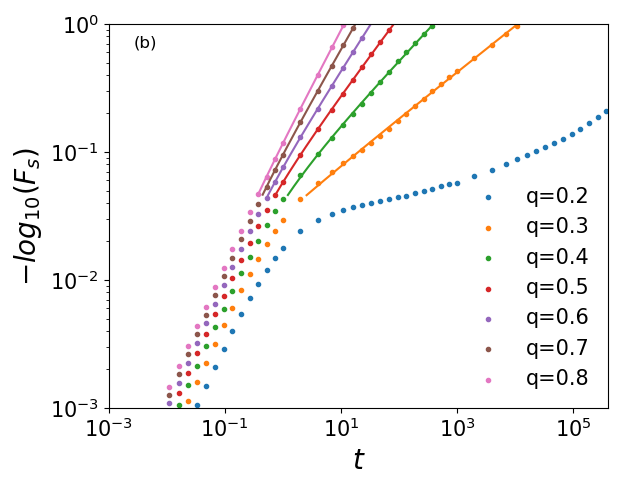}
  \caption{(a) Self-intermediate scattering function $F_s$ against time $t$ (symbols) in linear-log scales for various unblocking probability $q$ and density $\rho=0.8$.  (b) Same data as in (a) plotted as $-\log(F_s)$ versus $t$ in log-log scales.
(a)-(b) Data is fitted to the Kohlrausch-Williams-Watts (KWW) stretched exponential function for $F_s<0.9$ (lines). Data for $q=0.2$ is not fitted as system is only slightly relaxed.    
  }
  \label{fig_fs}
\end{figure}
We have measured the self-intermediate scattering function (SISF) defined as
\begin{equation}
  F_s(k,t) = \av{e^{ik \cdot (x_k(t)-x_k(0))}}
  \label{fs}
\end{equation}
for $k = (2\pi/\lambda)$ with $\lambda=10$ \cite{zhang2017}.
Results are shown in \fig{fig_fs}(a) for different $q$ and  $\rho=0.8$.
As expected of glassy systems, our data is well approximated by the Kohlrausch-Williams-Watts (KWW) stretched exponential function $A \exp[-(t/\tau)^{\beta}]$, where $A\simeq 1$, $\tau$ is the structural relaxation time and $\beta$ is the stretching exponent. \fig{fig_fs}(b) displays  a log-log plot of $-\log(F_s)$ against $t$. The approximately linear regions at large $t$ demonstrate the applicability of the KWW form.
\rev{As plotted in \fig{beta}, we observe that $\beta$ decreases from 0.86 to 0.37 as $q$ decreases from 0.8 to 0.3. Small values of $\beta$ in general indicate slowed dynamics expected of glass.
Results on SISF and $\beta$ at $\rho=0.2$ nevertheless exhibit more complicated properties and are reported in \App{rho02}.
}

\rev{Note that because the FRW is a defect model, SISF and $\beta$ calculated here may not be directly compared to MD measurements, which are typically particle based. In contrast,  this difference does not apply to the MSD. This is because under defect-induced motions, particle MSD only differs from defect MSD by a constant factor which equals the ratio between particle and defect populations \cite{gopinath2022}.}
  
\begin{figure}[tb]
  \includegraphics[width=0.9\columnwidth]{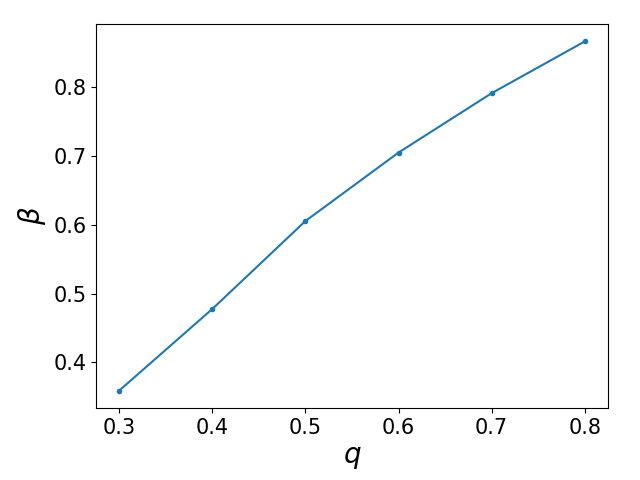}
  \caption{Stretching exponent $\beta$ against $q$ from KWW fits of the SISF in \fig{fig_fs}(a). }
  \label{beta}
\end{figure}


\hd{Diffusion coefficient}
\begin{figure}[tb]
  (a)\includegraphics[width=0.9\columnwidth]{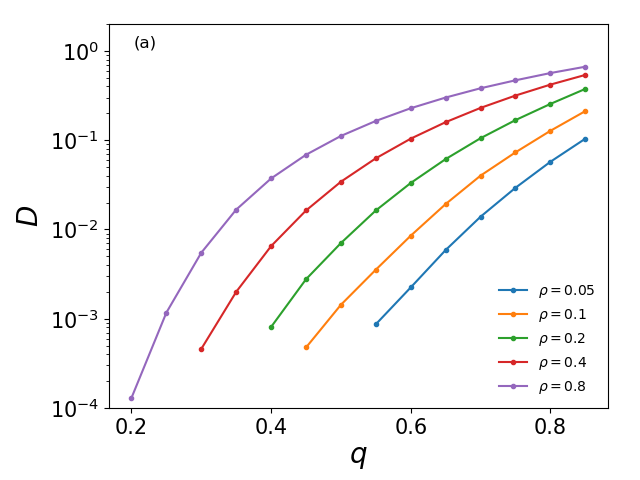}
  (b)\includegraphics[width=0.9\columnwidth]{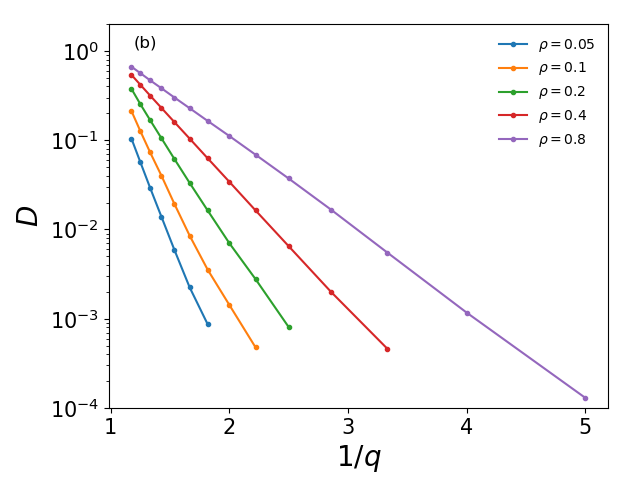}
  \caption{(a) Diffusion coefficient $D$  against unblocking probability $q$ (a) and $1/q$ (b) in semi-log scales for various density $\rho$.}
  \label{dq}
\end{figure}
We measure the MSD for various unblocking probability $q$ and density $\rho$, and calculate the particle diffusion coefficient as
\begin{equation}
  D=\frac{1}{2d}\lim_{t \to \infty}\frac{\text{MSD}}{t}.
  \label{D}
\end{equation}
We use the data regime where the slope in the log-log plot of MSD versus time is larger than 0.96, which we deem sufficiently close to 1.
\fig{dq}(a) shows $D$ against $q$ in a semi-log plot with different $\rho$.
As $q$ is expected to increase with temperature, we also plot $D$ against $1/q$ as shown in \fig{dq}(b). The near linear relation resembles an Arrhenius dependence of $D$ on $q$, noting that $q$ can be related to temperature as mentioned above. 

\section{Facilitation mechanism and ergodicity}
\label{facil}

Isolated particles are often trapped as show in \fig{traj}(a) for example. 
The maximum distance an isolated particle can travel from its initial position in either direction is bounded and follows a discrete exponential distribution with a mean
\begin{equation}
  \label{ltrap}
  s_{trap} = \frac{1}{1-q}-1.
\end{equation}
Including the initial site and noting the two possible directions of travel, the average trap size $W_{trap}$ of an isolated particle is
\begin{equation}
  \label{trapsize}
 W_{trap} =  \frac{2}{1-q} - 1
\end{equation}
which is finite for any $q<1$.

Therefore, isolated particles are either completely immobile or could only repetitively move within the trap. These motions are responsible for an initial rise in the MSD when they start to explore their traps. However, trap boundaries begin to limit their motions at longer time. This causes particles to hop repetitively within the trap, resulting in the MSD plateaus as shown in \fig{msd} for small $q$, corresponding to narrow traps.

In contrast, particles that are close to each other form mobile groups. For example, a mobile group consisting of 2 particles emerges at $q=0.7$ as shown \fig{traj_q}(a). At this value of $q$, all isolated particles are trapped, while coupled pairs are mostly mobile. The higher mobility of pairs compared to isolated particles clearly demonstrates facilitation in FRW.

The facilitation mechanism can be intuitively understood as follows. Assume that particle $k$ is bounded at time $t_1$ by a barrier bond $ij$ at an average distance $s_{trap}$, i.e. $w_{ijk}(t_1)=0$. The rate is quenched only until it is resampled by having the barrier broken. This is done by having another particle $l$ hop across the bond $ij$ at time $t_2>t_1$, which triggers a resampling with an unblocking probability $q$. If this results at $w_{ijk}(t_2)=w_0$, the barrier is lifted and particle $k$ is untrapped.
Conversely, particle $k$ can also untrap particle $l$. In general, if the individual traps of particles $k$ and $l$ overlap spatially, the pair may be able to continuously untrap each other, resulting at a mobile pair as observed in \fig{traj_q}(a).

As $q$ decreases, particle pairs become trapped permanently due to the more abundant constraints as observed in \fig{traj_q}(b), although the trap size of a pair is in general wider than that of a single particle. Nevertheless, adding a third particle enhances untrapping and hence facilitation, resulting in a mobile triplets.
Mobile triplets thus dominate motions as observed in \fig{traj_q}(b). 
Similarly, as seen in \fig{traj_q}(c), at an even smaller $q$, even triplets are immobile but there are mobile groups each of 4 coupled particles.

For any unblocking probability $q$, our results support that there is a dominant mobile group size of $m^*$ coupled particles. Smaller groups are immobile while larger groups are less abundant and play a lesser role in the dynamics. As $q$ decreases so that constraints are more numerous, $m^*$ increases because stronger facilitation with more frequent rate resampling is needed for  mobility. There is an associated sequence of mobility transitions as larger and larger groups become immobile. A quantitative study of these mobility transitions will be reported elsewhere.

Mobile groups, once existing in a FRW system, survive permanently. 
They may merge and split
as observed in Figs. \ref{traj} and \ref{traj_q} but do not vanish completely. This is because if such a system could turn into a trapped state, a trapped state can also evolve into a mobile one by the time reversal symmetry guaranteed by the detailed balance conditions. This contradict the assumption of a trapped state and hence cannot happen.

Ergodicity of large FRW systems is assumed when deriving equilibrium statistics in \sect{equil} and this will now be justified self-consistently.
We have explained in \sect{equil} that particles are randomly distributed with uniform probability at equilibrium, assuming ergodicity. Due to the random distribution, for any $q>0$ and $\rho>0$, a few sites with $m^*$ particles must exist initially in a sufficiently large system, where $m^*$ is the dominant mobile group size. These groups are mostly mobile and thus some of them must be able 
to traverse over the whole lattice at long time, untrapping, exchanging with and displacing all particles repeatedly. This justifies the ergodicity assumption.

\section{Discussions}
\label{conclusion}

\hd{Emergent facilitation}
Neither the facilitation process nor the dominant mobile group size in FRW is imposed directly by any rule in the model definition. The motions of particle $l$ can facilitate the motions of particle $k$ by lifting constraints as explained in \sect{facil}. However, it can equally likely inhibit the motions by installing new constraints.
Importantly, as all hops are reversible, inhibitions are only temporary and particles can wait until new constraints are reversed before advancing again. They impose no new permanent confinements and thus only a minor slowdown. In contrast, facilitation  successively opens up newer configurations, resulting at mobility. Therefore, facilitation is the emergent predominant impact of rate resampling, rather than inhibition. 

The FRW model is energetically trivial, like most KCM \cite{garrahan2011review}. However, we are not claiming that energetic interactions are unimportant in glass.  Kinetic constraints, i.e. $w_{ijk}=0$, occurring with probability $1-q$ according to \eq{wqt} model energetically unfavorable particle hops.
The FRW assumes that once all barriers with $w_{ijk}=0$ are identified, the constrained dynamics along the energy valleys can essentially be considered, as a first approximation, a flat potential energy landscape (PEL). This is sufficient for reproducing kinetic slowdown in glass.
However, thermodynamic quantities such as heat capacity cannot be studied in the current energetically trivial form of the FRW model, a situation similar to that of typical KCMs \cite{biroli2005,keys2013}.

\hd{Relation to the DPLM}
The DPLM is a particle model of glass with void-induced dynamics \cite{zhang2017}. Atoms are explicitly simulated, while voids are modeled effectively by empty sites. Interesting  physics of glass can be reproduced at low temperatures and low densities of voids
\cite{lulli2020,lee2020,lulli2021,lee2021,gao2022,gopinath2022,gao2023,li2023,ong2024,zhai2024}. It can thus be conceptually simpler and computationally faster if we simulate only the voids explicitly in an effective medium of particles. This has motivated us to define the FRW as a defect model in which particles corresponds to voids in the DPLM while atoms are not simulated explicitly. In the DPLM with energetic particle pair interactions, the motion of a void alters the local particle pairings and hence the PEL experienced by other voids. This is modeled in the FRW by the hopping rate resampling rule. On the other hand, the rate restoration rule in the FRW models the recovery of the system energy in the DPLM when particle configurations are reversed. These mechanisms in the DPLM have been motivated in turn by string interaction and repetition phenomena observed in our MD simulations \cite{lam2017}.

As a coarse-grained version of the DPLM, the FRW can be simulated much more efficiently. In particular, it can be simulated very efficiently in 1D, which is not naturally possible for the  DPLM because void-induced motions do not swap particles in 1D. Dynamics demonstrated by the FRW in general are qualitatively similar to those of the DPLM. \rev{In particular, the FRW exhibits a diffusion coeffcient power
law relating the diffusion coeffcient $D$ and the particle density $\rho$, fully consistent with that in the DPLM, which will
be explained elsewhere.}
%

The DPLM has been tested extensively against many glassy phenomena \cite{lulli2020,lee2020,lulli2021,lee2021,gao2022,gopinath2022,gao2023,li2023,ong2024,zhai2024}. The FRW may inherit many of these properties.
Nevertheless, the present FRW does not incorporate any particle interaction energy, which dominates the thermodynamics of the DPLM.
\rev{As a kinetic model,  we expect that the FRW should be able to reproduce equilibrium dynamical properties already demonstrated by the DPLM, such as surface enhanced mobility in glassy films \cite{zhai2024}. In addition, structural relaxation spectra \cite{yu2017,yu2024} can also be studied.}

\rev{It is straightforward to attain a thermodynamic variant of the FRW by a generalization to include an internal energy field dependent on the bond states  and this will be studied in the future. Then, it is promising to study many other glassy phenomena, especially those exhibited by the DPLM, including Kavacs paradox \cite{lulli2020} and effect \cite{lulli2021}, a wide range of kinetic fragility \cite{lee2020}, large heat capacity overshoot in fragile glass \cite{lee2021}, characteristic heat capacity of two-level systems at very low temperature \cite{gao2022}, Kauzmann paradox \cite{gao2023}, and fragile-to-strong transition \cite{ong2024}. 
Nevertheless, due to simplifications in the FRW, certain features such as diffusion-coefficient power laws upon partial swap of particles, a phenomenon demonstrated in both MD and DPLM simulations \cite{gopinath2022}, may not be reproduced because individual particles rather than defects must be explicitly considered.}

Another advantage of the highly simplified FRW is the possibility of an analytic description.
We have reported a local random configuration tree theory of glass which has been tested on the DPLM \cite{lam2018tree,deng2019}. The theory in fact has been motivated by analyzing the FRW. Essential assumptions in the theory such as a tree topology of the configuration space and a bimodal distribution of hopping rates are exact in the FRW by construction, while they hold approximately for the DPLM.
In addition, stringlike motions are assumed to strongly perturb the local energy landscape, resulting at resampling of the rates of all other strings in a local region.
The best fit to DPLM simulation data in 2D is obtained at a width $\sqrt{\mathcal V} = \sqrt{12} \simeq 3.5$ of these local regions \cite{lam2018tree}.
\rev{This means that within a $3.5 \times 3.5$ local region on a 2D DPLM lattice, voids fully facilitate each others' motions by being able to perturb each others' hopping rates. Using this as well as another parameter concerning energy fluctuations and after an analysis of the evolution of the system configuration in the local region, particle diffusion coefficient in the DPLM can be reasonably fitted over a wide temperature range \cite{lam2018tree}.}
We thus suggest that a site in the FRW roughly corresponds to a region of $3.5$ atoms wide.
We will apply the theory to describe the FRW quantitatively in the future for further testing and improvement of the theory.

\hd{Relation to other lattice models of glass}
The FRW and hence also the DPLM most closely resembles the FA model. In the $n$-spin facilitation variant of the FA model, one assumes a facilitation rule that $n$ neighboring spins must be present before they are allowed to flip \cite{fredrickson1984}.
\rev{The facilitation behavior as well as the parameter $n$ are directly imposed. 
They are analogous to the emergent property that $m^*$ particles in the FRW must be nearby to each other to form a mobile group, i.e. $n \sim m^*$.
The FA model may thus be considered as a simplified analogue of the FRW, taking key emergent facilitation behaviors in the FRW directly as defining rules.
Furthermore, the facilitation behaviors in the FRW emerge from the more fundamental hopping rate resampling and restoration rules. These directly imposed FRW rules are in turn analogous to emergent behaviors in the DPLM coming from the even more fundamental particle-dependent pair interactions \cite{zhang2017}.
From the perspective of the parameters, while $n$ in the FA model is directly chosen, its counterpart $m^*$ in FRW is an emergent value depending on the unblocking probability $q$. In addition, while $q$ in FRW is directly chosen, its counterpart in the DPLM can be estimated as an emergent value depending on the temperature \cite{lam2018tree}.  
Therefore, by comparing the FA model, the FRW and the DPLM, we can better understand intuitively how rules in the FA model may emerge naturally from more fundamental and  natural  physical considerations.}


Furthermore, most KCM including the FA model are energetically trivial  models. We believe that the relation among the FA model, the FRW and the DPLM explained above has a strong   implication to the study of KCM in general. 
Given a KCM, it should be interesting to search for a compatible energetic analogue so the origin and the thermodynamic consequence of the specific kinetic rules in the KCM can be  better understood.

The many interesting lattice models of glass \cite{garrahan2011review, fredrickson1984,zhang2017,tahaei2023,hasyim2024,nishikawa2024} should not only be applied to illustrate one or two facets of glassy properties, but ideally should be compatible with most, if not all, of the non-vibrational properties of glass.
Further tests of an individual model against a diverse range of glassy phenomena are urgently needed.

In conclusion, we have proposed the FRW as a simple KCM with random constraints. It shows emergent facilitation behaviors on the motions of neighboring particles. Dynamics is then dominated by mobile groups of size dependent on the assumed constraint density, analogous to facilitation rules in the seminal FA model of glass. The FRW is a defect  model and is a coarsened version of the atomistic DPLM. We believe that the study of compatible models  at different coarse-graining levels allows improved understanding of their applicability and limitation for the study of glass.

\acknowledgements

This work was supported by General Research Fund of Hong Kong (Grants 15210622 and 15303220).
~

\appendix

\section{System history encoded in fictitious states}
\label{states}
The instantaneous state of a FRW system is specified by the set $\{i_k\}$ of positions of all particles $k$. Nevertheless, due to the quenched randomness, a system simulated up to time $t$ can only be fully characterized by its complete history, i.e. full sets of positions $\{i_k(t)\}$  and hopping rates $\{w_{ijk} (t)\}$ for the whole duration from time 0 to $t$. Only then, hopping rate restoration can be possible. While a direct software implementation is possible, the data structures and the data retrieval algorithms are complex. The memory consumption is also huge.

Instead, we take the system history into account easily by effectively encoding it into expanded instantaneous system states. In this section, we introduce the general mathematical requirements involved. In the next section, our particular implementation will be explained.
To simplify the notation, we omit writing the $t$ dependence of the variables explicitly in the following.

For any particle $k$, we introduce a fictitious particle internal state $\Psi_k$. The bond between any NN sites $i$ and $j$ also admits a fictitious internal state $\Sij$ following $\Sij=\S_{ji}$. The expanded instantaneous state of the system is thus specified by the sets $\{i_k\}$, $\{\Psi_k\}$ and $\{\S_{ij}\}$. 
The hopping rate $w_{ijk}$ is deterministically calculated from
\begin{equation}
  \label{wPsiPhi}
     w_{ijk} = w(\Psi_k,\Sij) 
\end{equation}
where the function $w$ must be chosen to provide the required statistics followed by $w_{ijk}$.

After a hop of particle $k$ from site $i$ to site $j$, states are updated according to 
\begin{equation}  \label{F}
\begin{split}
\Psi_k' &= F(\xi_{ij},\Psi_k,\Sij) \\
\Sij' &= G(\xi_{ij},\Psi_k,\Sij)
\end{split}
\end{equation}
where
\begin{equation}
  \xi_{ij} =
  \begin{dcases}
    ~~1 & \text{for site $j$ on the right of site $i$}, \\
    -1  & \text{otherwise.}
  \end{dcases}
\end{equation}
The functions $F$ and $G$ must satisfy two conditions.
First,
to satisfy detailed balance, the new states must satisfy
\begin{equation}
\label{wsym}
  w(\Psi_k,\Sij) = w(\Psi_k',\Sij').
\end{equation}
Second,
a reversed hop of particle $k$ from $j$ to $i$ at the new states $\Psi_k'$ and $\Sij'$ must restore the original states,
\begin{equation}  \label{Fr}
\begin{split}
\Psi_k &= F(\xi_{ji},\Psi_k',\Sij') \\
\Sij &= G(\xi_{ji},\Psi_k',\Sij').
\end{split}
\end{equation}
Then, \eq{wPsiPhi} implies the restoration of also the previous rate $w_{ijk}$.
More generally, it is easy to see that a previous hopping rate can be restored even after an arbitrary sequence of related hops have occurred and reversed by applying \eqs{F}{Fr} recursively. Unrelated hops occurring at a distance, which do not affect these local states under consideration, naturally play no role in the reversal.  

As rate restoration is possible by straightforward calculations based on the instantaneous system state, memorization of previous rates and the system history is not required. The dynamics of FRW with the expanded states is thus reduced to simple Markovian processes for which standard kinetic Monte Carlo approaches apply.

\begin{figure}[tb]
  \includegraphics[width=0.9\columnwidth]{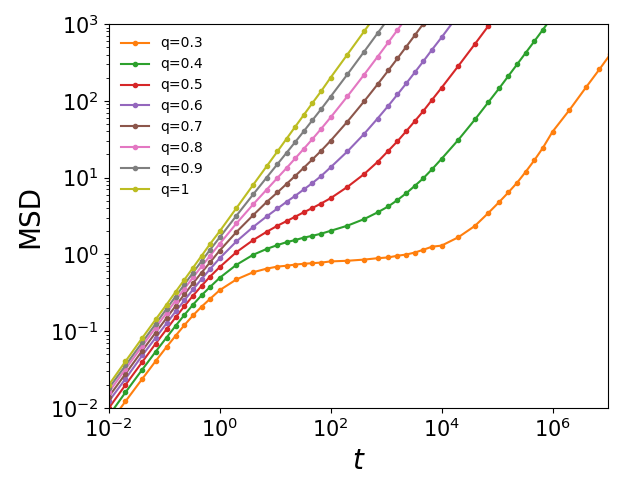}
  \caption{Particle MSD against time $t$ in log-log scales for various unblocking probability $q$ and density $\rho = 0.2$.}
  \label{rho02msd}
\end{figure}

\section{Rate restoration by reversible random numbers}
\label{algo}

We now explain our particular implementation of the expanded FRW states and their dynamics. To enable exact restoration of hopping rates, we adopt integer rather than float-point mathematics. As pseudo random rates with controllable statistics are required, we take note of the approach of congruential random number generators and consider only integers $0,1,\dots, \M-1$ following modulo-$\M$ arithmetics, i.e. $i \equiv (i \mod \M)$.  A natural choice is $\M = 2^{32}$, leading to 32-bits unsigned integer data type intrinsically available in most computers. This $\M$ is also sufficiently large to enable excellent pseudo random number properties.

Our algorithm requires non-trivial matrix inversions, implying that the states $\Psi_k$ and $\Sij$ cannot be simple integers. Instead, the next simplest choice is two-component matrices, i.e.
\begin{equation}
  \Psi_k = 
    \begin{pmatrix}
   \psi_k^{(1)} \\
   \psi_k^{(2)}
 \end{pmatrix},
 ~ ~ ~
 ~ ~ ~
  \S_{ij} = 
    \begin{pmatrix}
   \phi_{ij}^{(1)} \\
   \phi_{ij}^{(2)} 
  \end{pmatrix}
\end{equation}
where the elements $\psi_k^{(1)}$, $\psi_k^{(2)}$, $\phi_{ij}^{(1)}$  and $\phi_{ij}^{(2)}$ are 32-bits unsigned integers. Each $\Psi_k$ or $\S_{ij}$ is thus a 64-bit state.

\Eq{F}, concerning a hop of particle $k$ from site $i$ to site $j$,  is implemented as
\begin{equation}
  \label{FU}
\begin{split}
\Psi_k' &= [ U^{\xi_{ij}}(\Psi_k^T\Sij) ]^T \Psi_k \\
\Sij' &=  U^{-\xi_{ij}}(\Psi_k^T\Sij) \Sij
\end{split}
\end{equation}
where the superscript $T$ denotes transpose. A simple choice of $U(n)$ we have taken is
\begin{equation}
  \label{U}
U(n) =
  \begin{pmatrix}
  n+1 & n \\
    1 & 1
  \end{pmatrix}
\end{equation}
with an inverse
\begin{equation}
  \label{Uinv}
U^{-1}(n)=
  \begin{pmatrix}
   1 & -n \\
  -1 & n+1
  \end{pmatrix}.
\end{equation}
Modulo-$\M$ unsigned integer arithmetic is assumed in \eqr{FU}{Uinv} so that 
$-1\equiv \M-1$. The equations constitute congruential random number generators so that the elements $\psi_k^{(1)}$, $\psi_k^{(2)}$, $\phi_{ij}^{(1)}$  and $\phi_{ij}^{(2)}$ generated by iterating \eq{FU} in FRW simulations follow a uniform distribution in $0,1,\dots, \M-1$. This uniform distribution is readily verified in our simulations.

\Eq{FU} implicitly defines our choices of the functions $F$ and $G$. For any hop, the states $\Psi_k$ and $\Sij$ are updated via $U(n)$ and $U^{-1}(n)$ or $U^{-1}(n)$ and $U(n)$ respectively, so that the product
$\Psi_k^T\Sij$ is invariant after a hop, i.e.
\begin{equation}
  \label{PsiPhi}
  \Psi_{k}'^{T}\Sij' = \Psi_{k}^{T}\Sij.
\end{equation}
Similarly, forward and backward hops involve $U(n)$ and $U^{-1}(n)$ with the same $n$  which cancel out each other upon multiplication. Hence, states reversal, i.e. \eq{Fr}, is properly followed.

We define the rate function $w$ in \eq{wPsiPhi} as
\begin{equation}
  \label{w}
  w(\Psi_k,\Sij) = \w f(\Psi_k^T\Sij).
\end{equation}
where
\begin{equation}
  \label{f}
  f(n) = \theta( q - S(n)/\M )
\end{equation}
with $\theta$ being the Heaviside step function and $S(n)$ to be explained below.
As $\Psi_k^T\Sij$ is invariant during the hop (see \eq{PsiPhi}),  
the detailed balance condition (\ref{wsym}) is satisfied.

The function $S(n)$ can be the identity function, but we take a bitwise rotation operation by 16 bits to enhance randomization. Then, analogous to congruential generators, $\Psi_k^T\Sij$ and hence $S(\Psi_k^T\Sij)$ are also uniform random numbers in $0,1,\dots, \M-1$ so that $S(\Psi_k^T\Sij)/\M$ is a uniform random number in $[0,1)$. Therefore, $S(\Psi_k^T\Sij)>q$ occur with a probability $q$, implying that $w_{ijk}$ follows the required statistics according to Eqs. (\ref{wPsiPhi}), (\ref{w}), and (\ref{f}).

\begin{figure}[b]
  \includegraphics[width=0.9\columnwidth]{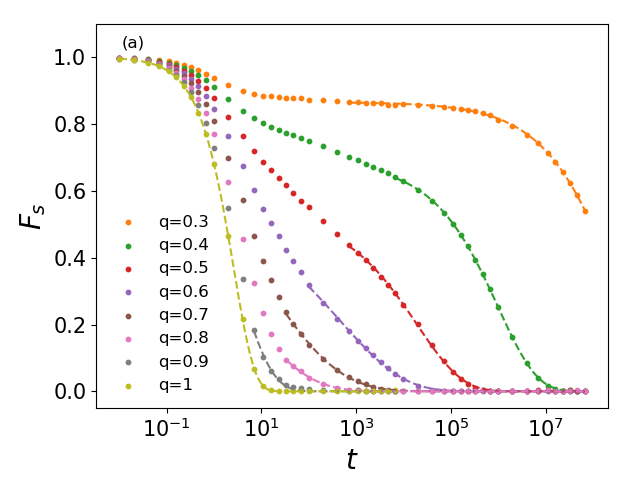}
  \includegraphics[width=0.9\columnwidth]{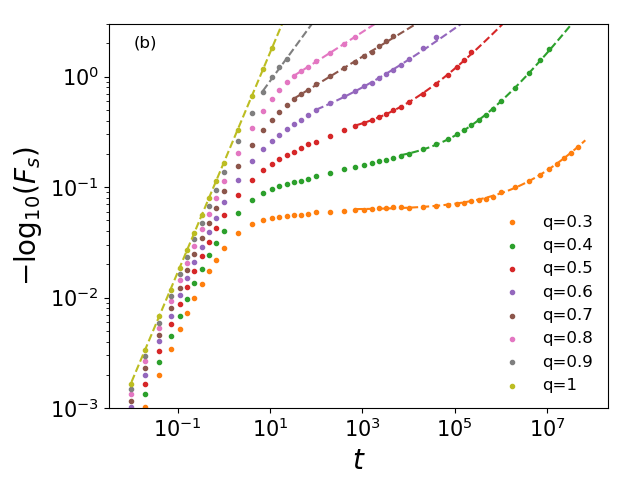}
  \caption{\rev{(a) Self-intermediate scattering function $F_s$ against time $t$ (symbols) in linear-log scales for various unblocking probability $q$ and density $\rho=0.2$.  (b) Same data as in (a) plotted as $-\log(F_s)$ versus $t$ in log-log scales.
(a)-(b) Data for the second relaxation step is fitted to the Kohlrausch-Williams-Watts (KWW) stretched exponential function (dashed lines).    } 
  }
  \label{fig_rho02fs}
\end{figure}

\section{Relaxation dynamics at low particle density}
\label{rho02}

The MSD is investigated at a particle density of $\rho = 0.2$ and results are plotted against time $t$ in \fig{rho02msd}. Similar to the case of   $\rho = 0.8$ in \fig{msd}, a plateau characteristic of glassy dynamics emerges at small unblocking probability $q$. 

\begin{figure}[tb]
  \includegraphics[width=0.9\columnwidth]{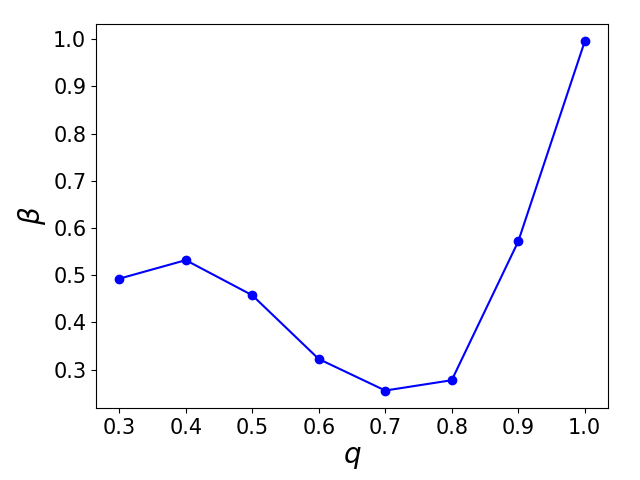}
  \caption{Stretching exponent $\beta$ against $q$ from KWW fits of the SISF in \fig{fig_rho02fs}. }
  \label{rho02beta}
\end{figure}

  We have also investigated the SISF, $F_s$, at $\rho=0.2$.
Figures \ref{fig_rho02fs}(a) and (b)  plot $F_s$ and $-\log(F_s)$ respectively.
Results show a two-step relaxation process at $q<1$, in contrast to a simple exponential relaxation at $q=1$, As $q$ decreases, the second step of the decay becomes more  dominant and fits reasonably well to the KWW stretched exponential function $A \exp[-(t/\tau)^{\beta}]$, where $A\le 1$, $\tau$ is the structural relaxation time and $\beta$ is the stretching exponent. Results are qualitatively quite similar to those for $\rho=0.8$ in \fig{beta}. Difference are more noticeable when plotting
$\beta$ against $q$ for $\rho=0.2$ in \fig{rho02beta}, showing a non-monotonic dependence on $q$. This is in contrast to the monotonic decrease of $\beta$ as $q$ decreases for $\rho=0.8$ in \fig{beta}. Nevertheless, a plateau, signifying glassy slowdown, emerges as $q$ decreases in both cases.

\bibliography{glass_short}

\end{document}